# FAST INSTABILITY CAUSED BY ELECTRON CLOUD IN COMBINED FUNCTION MAGNETS

S. A. Antipov, The University of Chicago, Chicago, IL 60637, USA
P. Adamson, A. Burov, S. Nagaitsev, M.-J. Yang, Fermilab, Batavia, IL 60510, USA

*Abstract*

One of the factors which may limit the intensity in the Fermilab Recycler is a fast transverse instability. It develops within a hundred turns and, in certain conditions, may lead to a beam loss. The high rate of the instability suggest that its cause is electron cloud. We studied the phenomena by observing the dynamics of stable and unstable beam, simulating numerically the build-up of the electron cloud, and developed an analytical model of an electron cloud driven instability with the electrons trapped in combined function dipoles. We found that beam motion can be stabilized by a clearing bunch, which confirms the electron cloud nature of the instability. The clearing suggest electron cloud trapping in Recycler combined function magnets. Numerical simulations show that up to 1% of the particles can be trapped by the magnetic field. Since the process of electron cloud build-up is exponential, once trapped this amount of electrons significantly increases the density of the cloud on the next revolution. In a Recycler combined function dipole this multi-turn accumulation allows the electron cloud reaching final intensities orders of magnitude greater than in a pure dipole. The estimated resulting instability growth rate of about 30 revolutions and the mode frequency of 0.4 MHz are consistent with experimental observations and agree with the simulation in the PEI code. The created instability model allows investigating the beam stability for the future intensity upgrades.

## FAST INSTABILITY

In 2014 a fast transverse instability was observed in the proton beam of the Fermilab Recycler. The instability acts only in the horizontal plane and typically develops in about 20 revolutions. The instability also has the unusual feature of selectively impacting the only first batch above the threshold intensity of $\sim 4 \times 10^{10}$ protons per bunch (Fig. 1). These peculiar features suggest that a possible cause of the instability is electron cloud. Earlier studies by Eldred et. al. [1] indicated the presence of electron cloud in the Recycler. The possibility of its trapping in the Recycler combined function magnets was suggested V. Lebedev [2].

The fast instability seems to be severe only during the start-up phase after a shutdown, with significant reduction being observed after beam pipe conditioning during the operation [3]. It does not limit the current operation with slip-stacking up 700 kW of beam power, but may pose a challenge for a future PIP-II intensity upgrade [4].

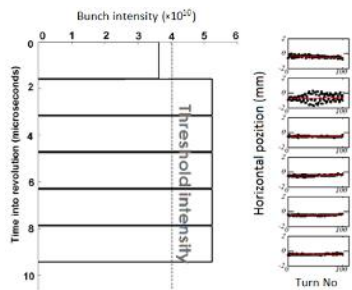

Figure 1: The first batch above the threshold intensity suffers the blow-up after injection into the ring [3].

## ELECTRON CLOUD TRAPPING

The most likely candidates for the source of electron cloud in Recycler are its combined function magnets. They occupy about 50% of the ring's circumference. In a combined function dipole the electrons of the cloud move along the vertical field lines, but the gradient of the field creates a condition for 'magnetic mirror' (Fig. 2) – an electron will reflect back at the point of maximum magnetic field $B_{max}$ if the angle between electron's velocity and the normal to the field lines is less than:

$$\theta_{max} = \cos^{-1}(\sqrt{B_0 / B_{max}}). \quad (1)$$

The particles with angles $\theta < \theta_{max}$ are trapped by magnetic field. For Recycler magnets (Table 1), Eq. (1) gives a capture of $\sim 10^{-2}$ particles of electron cloud, assuming uniform distribution.

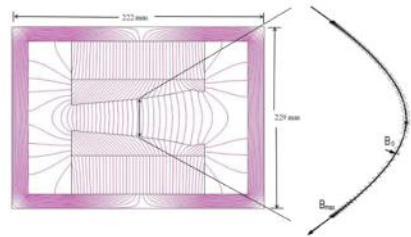

Figure 2: Electron cloud can get trapped by magnetic field of a combined function magnet. The picture shows the field lines inside a Recycler permanent CF dipole; the vacuum chamber is not shown.

Let us look at the process of electron cloud trapping in more detail. Let's assume that a proton bunch train has created some electron cloud and consider the last two bunches of the train. The first bunch kicks the electrons of the cloud, typically supplying them an energy of the order of a hundred eV. The electrons drift along the magnetic field in the vacuum chamber, finally reaching its walls and producing secondary electrons with the energies of a few eV [5]. In the absence of the beam these secondary electrons will eventually reach the aperture and die. But the next proton bunch can stop a fraction of the secondaries, reducing their angle $\chi$ to below the critical value (1). These electrons will remain trapped in the magnetic field after the beam is gone.

The presence of the second bunch is necessary to stop the particles, created by the first one, and therefore a single bunch cannot trap the cloud. Instead it clears the space, kicking the cloud to the physical aperture. The secondary electrons, created in the process, will eventually reach the vacuum chamber and be absorbed. This clearing bunch can be used to indicate the presence of the electron cloud [6] or to bring the electron cloud density below the threshold, stabilizing the beam.

The long-term confinement of the trapped electron cloud can be affected by two effects: longitudinal drift and scattering. The drift is caused by the horizontal derivative of the magnetic field $B' = dB/dx$. The longitudinal drift velocity is

$$v_d = \frac{1}{2}\omega_c r_c^2 \frac{B'}{B_0}, \quad (2)$$

where $\omega_c = eB_0/(m_e c)$ is the cyclotron frequency, $B_0$ – the dipole magnetic field component, and $r_c$ – the radius of the orbit. If the cloud drifts a distance $l_d$ comparable with the magnet length $L_{dip}$ it may escape the magnet and decay. For the Fermilab Recycler $v_d < 2 \times 10^5$ cm/s, $L_{dip} = 5$ m and the lifetime is ~ 1 ms.

In general, the lifetime of trapped electrons may be also limited by scattering on each other and on the residual gas. For the scattering on the other electron cloud particles the Coulomb cross-section $\sigma_C$ can be estimated as (see Eq. (41.7) in [7])

$$\sigma_C = 16\pi r_e^2 \left(\frac{m_e c^2}{\varepsilon}\right)^2 \Lambda_C, \quad (3)$$
$$\Lambda_C = \ln(1/\chi_{min})$$

where $r_e$ – is the classical electron radius, $\varepsilon$ – the electron energy, $c$ – the speed of light, $\Lambda_C$ – the Coulomb logarithm, and $\chi_{min}$ – the minimal scattering angle. $\chi_{min}$ can be estimated as $\chi_{min} \sim \chi_{max}$, since the scattering does not lead to a particle loss if it stays within the trapping cone $\chi < \chi_{max}$. Then for the electron energies $\varepsilon \sim 1-10$ eV the cross-section $\sigma_C < 10^{-13}$ cm$^2$.

According to the experimental measurements [8], the scattering cross-section for many residual gases is of the order of $10^{-15}$ cm$^2$ at the energies $\varepsilon < 10$ eV. Combining the two scattering effects we obtain a lifetime ~ 1 ms for the electron cloud density $n_e < 10^7$ cm$^{-3}$ and the residual gas pressure $P \sim 10^{-8}$ Torr.

Since all the loss mechanisms result in the lifetime much larger than the revolution period of 11 μs, all the trapped cloud will be present on the next turn. It will act as the new seed electrons, and can lead to a higher electron cloud density on the next revolution.

## NUMERICAL SIMULATION

We simulated electron cloud build-up over multiple revolutions in a Recycler dipole using the PEI code [9]. The code simulates the build-up and 2D transverse coupled motion of the electron cloud and the beam. The electron cloud is represented by an ensemble of macroparticles of a constant weight, and the beam – by a series of rigid bunches with Gaussian transverse shape. The beam-cloud interaction is calculated using the Basetti-Erskine model [10]. The ring was modelled as a linear transfer matrix with one interaction point, representing a combined function magnet. The input parameters of the simulation are summarized in Table 1.

The main source of primary electrons in Recycler is the collisional ionization of residual gas by the beam. To simulate it we put the seed electrons at the beam center with their linear density following $\lambda$[m$^{-1}$] ~ $6N_b P$[Torr], where $N_b$ is the number of protons in a bunch [11].

The model of secondary emission includes true secondary and elastically reflected electrons and assumes normal incidence [12]. In a dipole field, however, an electron hits the wall of a vacuum chamber at an angle. That increases the time the electron spends near the wall surface and consequently the SEY. Experimental data on angular dependence of SEY fits an empirical formula:

$$SEY = SEY_0(1 + a_1(1 - \cos(\chi)^{a_2})), \quad (4)$$

where $a_1$ and $a_2$ are material specific parameters and $SEY_0$ – the yield measured at normal incidence [13]. For a simple estimate one can use $a_1 = 0.26, a_2 = 2$. Then for the Fermilab Recycler combined function dipoles the simulated mean incident angle is 15 deg (Fig. 3) and the resulting increase of SEY, according to Eq. (4) is about 5%.

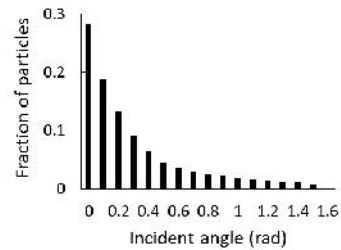

Figure 3: Most particles hit the vacuum chamber at small angles to the normal.

For a pure dipole field, the cloud rapidly builds up during the passage of the bunch train and then decays back to the initial ionization electron density in about 300 RF buckets, or ~ 6 μs (Fig. 4). When the field gradient is added, up to 1% of the electron cloud stays trapped, increasing the initial density on the next revolution. The final density, which the cloud reaches after ~ 10 revolutions, can be as high as two orders of magnitude greater than in the pure dipole case (Fig. 4). The resulting cloud distribution is a stripe along the magnetic field lines, with higher particle density being closer to the walls of the vacuum chamber (Fig. 5).

At lower densities ~$10^{-2}$ of particles are trapped, which agrees with an estimate from Eq. (1); as the density of electron cloud increases the trapping ratio goes down to ~$10^{-3}$, probably due to the space charge of electron cloud.

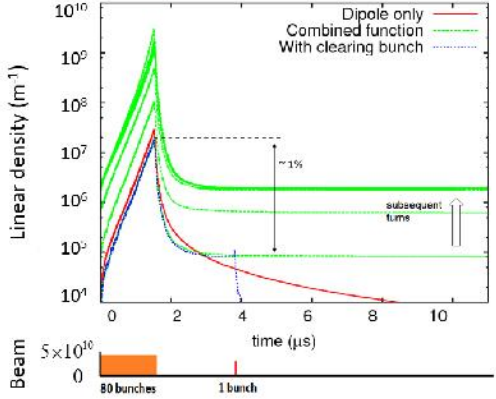

Figure 4: In a combined function magnet the electron cloud accumulates over many revolutions, reaching much higher line density, than in a pure dipole. A clearing bunch destroys the trapped cloud, preventing the accumulation.

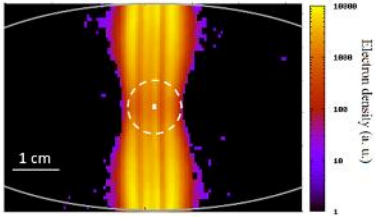

Figure 5: Electron cloud forms a stripe inside the vacuum chamber; beam center and its 2- boundary are shown in white.

Table 1: Recycler parameters for simulation in PEI

| | |
|---|---|
| Beam energy | 8 GeV |
| Machine circumference | 3.3 km |
| Batch structure | 80 bunches, 5e10 p |
| Tunes: x, y, z | 25.45, 24.40, 0.003 |
| RF harmonic number | 588 |
| RMS bunch size: x, y, z | 0.3, 0.3, 40 cm |
| Secondary emission yield | 2.1 at 250 eV |
| Density of ionization e$^-$ | $10^4$ m$^{-1}$ (at $10^{-8}$ Torr) |
| B-field and its gradient | 1.38 kG, 3.4 kG/m |
| Beampipe | Elliptical, 100 x 44 mm |

As mentioned above, a trapped cloud can be cleared by a single bunch following the beam at a sufficient distance. In Fig. 4 a bunch of $5\times10^{10}$ protons, added 120 RF buckets after the main batch, destroys the trapped cloud, preventing the multi-turn build-up. First, one can see a small increase in the cloud density as the clearing bunch kick the cloud and it reaches the vacuum chamber, producing the secondary electrons. Then, the density rapidly drops as these secondaries reach the aperture.

The multi-turn electron cloud accumulation due to the trapping mechanism might play an important role in a proton ring, where the density of the primary ionization electrons is relatively low. For a positron machine of a similar energy the amount of primary electrons is much greater due to the photoemission by synchrotron radiation. Because of the large number of primary electrons, the cloud can reach a saturation density during the passage of on bunch train. Then the presence of trapping would only slightly affect the overall picture, shifting the saturation towards the head of the batch. The recent studies at CESR show that, the cloud in its combined function magnets reaches a saturation density during the passage of one positron bunch train [14].

## WITNESS BUNCH TEST

We used a clearing bunch technique, similar to that used at Cornell [6] to check whether the instability is caused by trapped electron cloud. If a trapped electron cloud is present in the machine, a single bunch of high enough charge following the main batch, will kick it and clear the aperture. This clearing of electron cloud then can be noted by observing a change in beam dynamics.

Figure 6 (top) shows the increase of beam center oscillations, measured by BPMs, of an unstable batch just above the threshold intensity. The batch consists of 80 bunches of $4.6\times10^{10}$ p. The horizontal oscillations rapidly grow, leading to beam dilution and a loss of a fraction of intensity, then the beam is stabilized by the dampers. When a single clearing bunch of $\geq 1\times10^{10}$ p is injected in the machine before the high-intensity batch, the later remains stable (Fig. 6, bottom). The position of the clearing bunch does not change the picture – it can be as far as half of the ring (or ~ 5 μs) apart from the batch, suggesting that there is a portion of the electron cloud that survives over one revolution, and it can be removed with a clearing bunch. This agrees qualitatively with the simulation of electron cloud build-up and trapping in Recycler dipoles (Fig. 4).

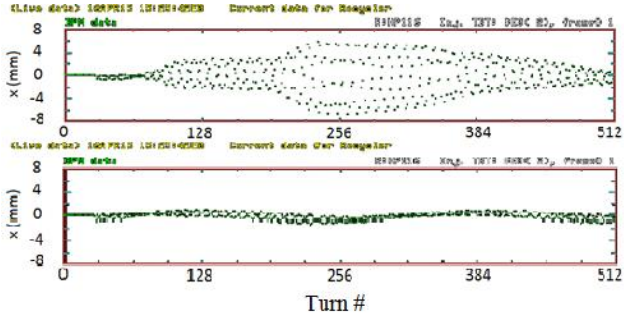

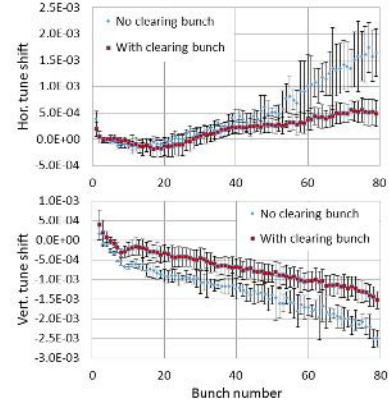

Figure 6: Without the clearing bunch the beam of $3.6 \times 10^{12}$ p blows up in about 20 turns (top); with the clearing bunch of $1 \times 10^{10}$ p it remains stable (bottom). Turn-by-turn measurement of the horizontal position of the beam center.

The presence of electron cloud provides additional focusing or defocusing, shifting the betatron tunes. Since the space charge does not change the coherent tune and the resistive wall creates a negative tune shift, a positive horizontal tune shift, if observed, would be a clear signature of the presence the electron cloud. The tune shifts were measured with a stripline detector. The detector consists of two horizontal and two vertical 1.4 m long plates (quaterwave for 53 MHz) with 50 wave impedance inside a round vacuum chamber (Fig. 7). The detector has a 1 GHz bandwidth and a linear response within 75% of the physical aperture of 110 mm [15].

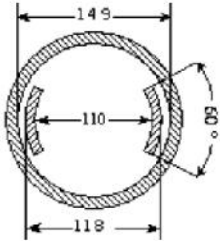

Figure 7: Cross section of Recycler stripline detector [15].

Figure 8 depicts the betatron tune shift within the 80-bunch train with respect to the first bunch, measured over 600 revolutions with a stripline detector, with the dampers off during the measurement. The positive horizontal tune shift is a clear signature of the presence of a negative charge at the beam center. The vertical tune shift is negative, indicating that the maximum density of the cloud is outside the beam, which agrees qualitatively with the simulated distribution (Fig. 5). When a clearing bunch is added, the tune shifts decrease, indicating a reduction of electron cloud density, which agrees with the simulation (Fig. 4). The remaining linear slope in the vertical tune shift is likely to be due to the resistive wall impedance. According to the recent measurements, in Recycler the vertical impedance is about five times larger than the horizontal [16].

Figure 8: Presence of the clearing bunch reduces the tune shift between the head and the tail of the high-intensity bunch train: $5*10^{10}$ ppb, 80 bunches. The error bars represent the spread between different measurements.

## SIMPLE ANALYTICAL MODEL OF BEAM-CLOUD INTERATION

First, consider a round coasting proton beam travelling in a ring, uniformly filled with electron cloud. Let us denote the position of the beam centroid at an azimuthal angle $\theta$ at time $t$ as $X_p(t, \theta)$. Further, assume that the beam travels at a constant azimuthal velocity around the ring $\omega_0$ and use a smooth focusing approximation with a betatron frequency $\omega_\beta$.

For simplicity, one can represent the electron cloud by a cylinder of a uniform charge density $n_e$ and the same radius as the proton beam, located at a horizontal position $X_e$. Let us further assume that the total number of electrons does not change in time. Because of the vertical dipole field, the individual electrons of the cloud cannot drift horizontally, but the position $X_e$ can change as some regions build up and others are depleted, following the transverse motion of the proton beam. The characteristic time constant of this slow motion of the electron cloud is then the time of build-up: $\eta \sim 1/\tau_{buildup}$.

For small oscillation amplitudes we can assume the electron-proton interaction force to be linear in displacement. Then the coupled collective motion of the beam and the electron cloud is described by the following system of equations:

$$\begin{cases} \left(\frac{\partial}{\partial t} + \omega_0 \frac{\partial}{\partial \theta}\right)^2 X_p + \Gamma\left(\frac{\partial}{\partial t} + \omega_0 \frac{\partial}{\partial \theta}\right) X_p = \\ \qquad = -\omega_\beta^2 X_p + \omega_p^2 (X_e - X_p), \quad (5) \\ \frac{\partial}{\partial t} X_e = \eta (X_p - X_e) \end{cases}$$

where $\Gamma$ is the rate of Landau damping. The coupling frequency $\omega_p$ is approximated as

$$\omega_p^2 = \frac{2f n_e r_p c^2}{\gamma}, \quad (6)$$

where $r_p$ is the classical proton radius and $\gamma$ – the relativistic factor.

The linear damping term $\Gamma$ in Eq. (5) arises from the spread in betatron frequencies for particles oscillating with different amplitudes. The characteristic rate of the Landau damping can be estimated as

$$\Gamma \sim \omega_s \frac{\Delta Q_x}{Q_x}, \quad (7)$$

where $Q_x$ is the horizontal tune and $\Delta Q_x$ is its rms spread.

Looking for solutions of Eq. (5) in a form $X_{e,p} \propto e^{-i\omega t + in\theta}$ one obtains an equation for the mode frequency $\omega$:

$$-(\omega - n\omega_0)^2 - i\Gamma(\omega - n\omega_0) + \omega_s^2 + \frac{\omega \omega_p^2}{\omega + i\nu} = 0 \quad (8)$$

It can be solved perturbatively, under the assumption that

$$\omega_s, \nu \gg \omega_0, \omega_p, \Gamma. \quad (9)$$

Solving Eq. (8) for each integer wave number $n$ one gets:

$$\omega = n\omega_0 + \omega_s + \Delta\omega, \quad (10)$$

where the small complex tune shift $|\Delta\omega| \ll \omega_s, \nu, \omega_s$ is:

$$\Delta\omega \approx \frac{1}{2}\left[-i\Gamma + \frac{\omega_p^2}{\omega_s} \frac{\omega(\omega - i\nu))}{\nu^2 + \omega^2}\right] \quad (11)$$

The imaginary tune shift in Eq. (11) consists of two parts with the first being the Landau damping term. The most unstable mode, for which $\text{Im}(\Delta\omega)$ is the greatest, is $\omega_{\max} = \nu$ and its wave number $n_{\max}$ is

$$n_{\max} = \frac{\nu - \omega_s}{\omega_0} = \frac{\nu}{\omega_0} - Q_x, \quad (12)$$

and the growth rate of this mode is

$$\lambda_{\max} = \frac{1}{2}\left(\frac{\omega_p^2}{2\omega_s} - \Gamma\right). \quad (13)$$

The threshold electron cloud density $n_{e,thr}$ can be found from the condition $\lambda_{\max} = 0$, which yields

$$n_{e,thr} = \frac{\pi \Gamma \omega_s}{\nu c^2 r_p}. \quad (14)$$

Since, in general, the strength of Landau damping $\Gamma$ depends on the density of the electron cloud, this equation might have one, many, or no solutions at all [17].

In an experiment an observer will see the most unstable mode as it suppresses the others thanks to its higher exponential growth rate. Thus, a turn-by-turn measurement of beam position at $\theta = \theta_0$ will detect a tune shift of

$$\Delta Q_{\max} \approx \frac{1}{4Q_x} \frac{\omega_p^2}{\omega_0^2}. \quad (15)$$

Knowing the complex frequency shift $\Delta\omega$ we can find the impedance of the cloud as (see for example [18] Eq. (6.262)):

$$Z(\omega) = Z_0 \frac{\pi T_0^2 \omega_s}{2\pi Nr_p} i(\Delta\omega + i\Gamma/2 - \omega_p^2/(2\omega_s)), \quad (16)$$

where $N$ is the number of protons in the ring and $Z_0$ is the vacuum impedance. Because the electron cloud shifts both the coherent and the incoherent frequencies, we subtracted here the incoherent tune shift.

Knowing the impedance one can compute the wake functions using the formula (2.72) from [18]:

$$W(z) = \frac{-2i}{Z_0 c} \int_{-\infty}^{+\infty} Z(\omega) e^{i\frac{\omega z}{c}} d\omega \quad (17)$$

In the case of a bunched beam, in a rigid bunch approximation, one needs to compute $W(z)$ only in a discreet set of bunch positions $z_k = kc\tau_{rf}$, where $\tau_{rf}$ is the RF period.

Finally, from the impedance of the most unstable mode one can estimate the instability growth rate of a bunched beam as [19]:

$$\lambda_{b,\max} \approx -\frac{L}{C} \frac{8\pi r_p N_b \omega_x}{\gamma \tau_{rf} c} \frac{\text{Re}(Z(\omega_{\max}))}{Z_0} - \frac{1}{2}\Gamma, \quad (18)$$

where $C$ is the ring circumference and $L$ is the total length of the magnets. For the Recycler $L/C \approx 1/2$.

## INSTABILITY IN RECYCLER

In order to use the model and estimate the parameters of the fast instability in Recycler one needs to know the density of the electron cloud and the rate of its build-up. We obtain these quantitative parameters by measuring the betatron frequency shift and comparing it with the build-up simulations.

We injected one batch of 80 proton bunches of $5 \times 10^{10}$ ppb and measured the shift of the horizontal tune as a function of bunch number. Because the positive horizontal tune shift is a distinctive feature of the electron cloud, it allowed us an estimation of the cloud density. In order to check with the simulation the cloud density both within the high-intensity batch and after its passage we put a witness bunch of low intensity – $8 \times 10^9$ p, insufficient to clear the electron cloud, at different positions behind the main batch.

The experimental results are in good agreement with the simulation (Fig. 9) and the small discrepancies may come from the multiple assumptions used in Eq. (15). The resulting dependence allows the estimation of the maximum density of electron cloud $n_e \sim 6 \times 10^{11} \text{m}^{-3}$. The density increases by an order of magnitude in 40 bunches (800 ns) and falls after the beam has passed in 10 bunches (200 ns). The characteristic rate of the build-up is about 1/20 bunches or $\nu \sim 2.65 \times 10^6 \text{s}^{-1}$. The parameters of the model are summarized in Table 2.

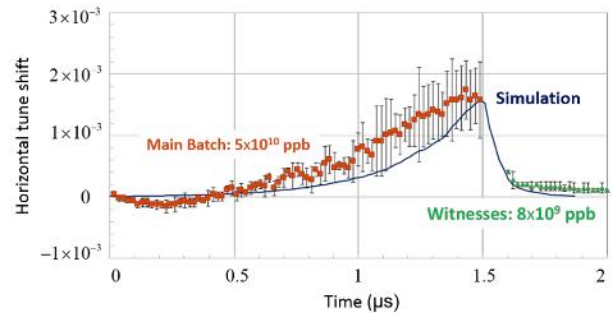

Figure 9: Results the of electron cloud simulation agree with the measured horizontal tune shift. Beam: $5 \times 10^{10}$

ppb, 80 bunches, followed by one witness bunch of $0.8 \times 10^{10}$ p at various positions. The gap between the high-intensity batch and the witness is due to the rise-time of the injection kickers.

The most unstable mode has the frequency of about 0.4 MHz, its impedance, calculated using Eq. (16), is 20 MΩ/m (Fig. 10). Figure 11 depicts the corresponding wake function $W(n)$ as a function of bunch number $n$. $W(n)$ fits an exponential decay curve

$$W = W_0 \exp(-\Delta z / c\lambda), \Delta z > 0. \qquad (19)$$

The estimate of the mode frequency qualitatively agrees with the simulation in the PEI code and the stripline measurement. PEI simulated the ring, completely filled with 588 bunches of $5 \times 10^{10}$ p. The resulting frequency is about 0.7 MHz (Fig. 12). In the stripline measurement one batch of 80 bunches of the same charge was injected. The measured frequency was about 0.9 MHz. Both simulated and measured frequencies agree on the order of magnitude with each other and the estimate.

Using the calculated value of the real part of the impedance we can now estimate the growth rate using Eq. (13). We obtain the growth rate of $\varkappa_{b,max} = 0.033$ and the characteristic time of the instability $\tau_{max} = 1/\varkappa_{b,max} \approx 30$ turns.

The threshold electron cloud density, calculated using Eq. (14), $n_{e,thr} = 8.2 \times 10^{10} \mathrm{m}^{-3}$. This density is achieved at the proton intensity of about $4.5 \times 10^{10}$ ppb, which is also consistent with experimental observations.

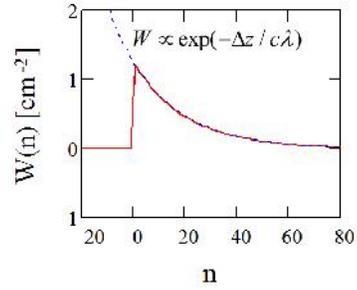

Figure 11: Electron cloud wake falls down exponentially with distance.

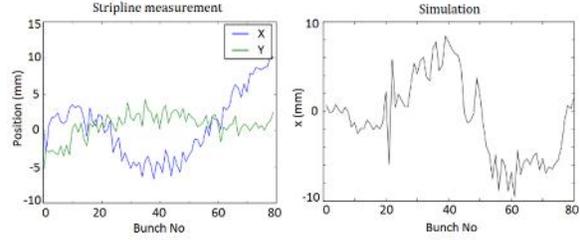

Figure 12: Simulation in PEI and stripline measurements show an instability in the horizontal plane with a period of slightly less than the length of a batch and a frequency < 1 MHz.

Table 2: Parameters of the model

| Parameter | Symbol | Value |
| --- | --- | --- |
| Relativistic factor | | 10 |
| Cyclotron frequency | $\omega_0$ | $0.57 \times 10^6$ s$^{-1}$ |
| Betatron frequency | $\omega_\beta$ | $14.54 \times 10^6$ s$^{-1}$ |
| Protons per bunch | $N_b$ | $5 \times 10^{10}$ |
| Electron cloud density | $n_e$ | $6 \times 10^{11}$ m$^{-3}$ |
| e-p coupling frequency | $\omega_p$ | $0.23 \times 10^6$ s$^{-1}$ |
| Build-up rate | $\lambda$ | $2.65 \times 10^6$ s$^{-1}$ |
| Chromatic tune spread | $\Delta Q_x$ | $2.7 \times 10^{-3}$ |

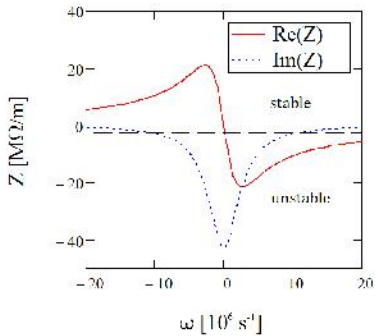

Figure 10: Real and imaginary parts of impedance as a function of a mode angular frequency $\omega$. The dashed line represents the Landau damping. The modes with Re($Z(\omega)$) below the dashed line are unstable.

## CONCLUSION

A fast transverse instability in the Fermilab Recycler might create a challenge for PIP-II intensities. Understanding its nature is important for making predictions about the machine performance at higher intensities.

We have observed that the fast instability can be mitigated by injection of a single low intensity clearing bunch. This finding suggests that the instability is caused by electron cloud and the cloud is trapped in Recycler magnets. Bunch-by-bunch measurements of the betatron tune have shown its shift towards the end of the bunch train. The tune shift decreases after the addition of the clearing bunch, which is also consistent with the electron cloud picture.

There is practically no doubt that the source of trapping is the combined function magnets, occupying around 50% of Recycler circumference. Combined function magnets are widely used in contemporary accelerators and are a technology of choice for some future machines. According to numerical simulations in PEI, ~$10^{-2}$–$10^{-3}$ of particles are trapped by magnetic field of those magnets. That allows the electron cloud to gradually build up over multiple turns, reaching final intensities orders of magnitude greater than in a pure dipole. The results of electron cloud build-up simulation in the Recycler combined function dipoles agree qualitatively with the observed stabilization of the beam by a clearing bunch and quantitatively with the measurements of betatron tune shift. According to the simulations, the estimated cloud density is $6 \times 10^{11}$ m$^{-3}$ on the beam axis and the characteristic times of its build-up and decay are 40 and 10 RF periods respectively.

We have created a simple analytical model of the transverse multibunch instability, driven by the electron cloud trapped inside the combined function magnets. The model allows an estimation of the instability threshold, the frequency of the most unstable mode and its growth rate. For the current parameters of the Recycler beam we find the mode with a frequency of 0.4 MHz and a growth rate of 30 revolutions, which is consistent with the observations of the fast instability and the simulations in PEI. The model allows the prediction of the rate of the instability for higher intensities of the proton beam, given an estimate of the electron cloud density, which can be obtained from numerical simulations.

## ACKNOWLEDGMENT

The authors are grateful to K. Ohmi (KEK) for his help with PEI code and to Yuri Alexakhin (FNAL) for multiple useful suggestions.

Fermilab is operated by Fermi Research Alliance, LLC under Contract No. DE-AC02-07CH11359 with the United States Department of Energy.

## REFERENCES

[1] J. Eldred *et al.*, in *Proc. HB'14,* pp. 419-427
[2] V. Lebedev, private communication
[3] J. Eldred, "The High-Power Recycler: Slip-stacking & Electron Cloud", Fermilab, Nov. 2015
[4] S. Holmes *et al.*, in *Proc. IPAC'16*, pp. 3982-3985
[5] G. Iadarola, "Electron Cloud Studies for CERN Particle Accelerators and simulation Code Development", Ph.D. thesis, Beams Dep., CERN, CERN-THESIS-2014-04, 2014
[6] M. Billing *et al.,* "Measurement of electron trapping in the Cornell Electron Storage Ring", *Phys. Rev. ST Accel. Beams* vol. 18, p. 041001, Apr. 2015
[7] L. D. Landau, E. M. Lifshiz, *Physical Kinetics*, Moscow: Fizmatlit, 2002
[8] A. D. MacDonald, *Microwave Breakdown in Gases*. New York, NY: Wiley, 1966
[9] K. Ohmi, "Electron cloud effects: codes and simulations at KEK", CERN Rep. CERN-2013-002, pp. 219-224
[10] M. Bassetti and G. A. Erskine, "Closed expression for the electrical field of a two-dimensional Gaussian charge", CERN-ISR-TH-80-06, CERN, 1980
[11] F. Zimmermann, "Particle-matter interaction" in *Handbook on Accelerator Physics*, A. Chao, M. Tigner, Ed. World Scientific, 1999
[12] Benedetto E. *et al.*, "Review and Comparison of Simulation Codes Modeling Electron-Cloud Build Up and Instabilities", in *Proc. EPAC'04*, Lucerne, Switzerland, 2004, pp. 2502-2504
[13] M. A. Furman, M. Pivi, Probabilistic model for the simulation of secondary electron emission, *Phys. Rev. ST Accel. Beams* vol. 5, p. 124404, 2002
[14] J. A. Crittenden, Y. Li, S. Poprocki, J. E. San Souice, "Electron cloud simulations for the low-emittance upgrade at the Cornell Electron Storage Ring", in *Proc. NAPAC'16*, Chicago, IL, Oct. 2016
[15] J. Crisp, K. Gubrienko, V. Seleznev, "Stripline Detectors for Main Injector", in *Proc. National 16th Conf. on Accelerators*, Protvino, Russia. Oct. 1998
[16] R. Ainsworth, P. Adamson, A. Burov, I. Kourbanis, M.-J. Yang, "Estimating the transverse impedance in the Fermilab Recycler", in *Proc. IPAC'16*, Busan, 2016, pp. 867-869
[17] A. Burov, "Three-beam instability in the LHC", FERMILAB-PUB-13-005-AD, 2013
[18] A. Chao, *Physics of Collective Beam Instabilities in High Energy Accelerators*, Ney York, NY: Wiley, 1993
[19] Yu. Alexahin, FNAL Beams-doc-4863-v1, Fermilab, Jun. 2015